\documentstyle[12pt]{article}

\begin{document}
\title{\textbf{Curvature and folding dynamo effects in turbulent plasmas and ABC flux tubes}} \maketitle
\author{L.C. Garcia de Andrade \\
\small Departamento de F\'{\i}sica Te\'orica-IF-UERJ-RJ, Brasil}
\maketitle
\vspace{0.1cm} \paragraph*{Investigation of the eigenvalue spectra of dynamo solutions, has been proved fundamental for the knowledge of dynamo physics. Earlier, curvature-folding relation on dynamos in Riemannian spaces has been investigated [PPL 2008]. Here, analytical solutions representing general turbulent dynamo filaments are obtained in resistive plasmas. Turbulent diffusivity with vanishing kinetic helicity yields a fast mode for a steady dynamo eigenvalue. The magnetic field lays down on a local frame 2 plane along the filaments embedded in a 3D plasma. Curvature effects plays the role of folding in fast magnetic dynamos. In the present examples, plasma equipartition between normal and binormal components of the magnetic field components is considered. In the opposite case, oscillatory, purely imaginary, branches of the spectrum are found in dynamo manifold. Degenerate eigenvalues, are obtained when the dynamo growth rate coincides with the filaments curvature. Spectra of dynamo obtained are similar to the fast dynamo solution obtained by Arnold on a compact torus. Dynamo experiments making use of this kind of solution have been investigated by Shukurov et al [Phys Rev E, 2008] with Perm liquid sodium experiments on a circular torus implications. Another example of dynamo plasma is given by the Arnold-Beltrami-Childress [ABC] twisted magnetic flux tubes with stagnation points. PACS: 47.65.Md. Key-word: dynamo plasma.}
\newpage
\section{Introduction}
 The simplest fast dynamo solution of a self induction equation of a stationary flow undergoing uniform stretching has been given by Arnold et al \cite{1}. Following this solution up to the present date many numerical simulations have been obtained \cite{2,3}. More recently Garcia de Andrade has investigated \cite{4,5} the relation between the dynamo compact manifold Riemann curvature and the folding of the stretch-twist-fold dynamo mechanism, in realm os solar physics. Also recently Nunez \cite{6} has investigated the role of stretching also in the case of the back reaction dynamos, instead of the pure kinematic dynamo effects. In his case, the local dynamics is governed by the decrease of the growth rate of the magnetic field, with minus the square of the eigenvalue. This definetely shows that investigation of the eigenvalue spectra in dynamo manifolds is fundamental for the better comprehension of the dynamo physics. In this paper, new analytical solutions of filamentary turbulent fast dynamos \cite{7} corresponding to a Anosov magnetic dynamos. Anosov dynamos are basically Arnold dynamos which develope stretching along the eigendirections of the magnetic fields in dynamo flow. The corresponding stretching and squeezing eigenvalues are slightly different from the ones obtained by Arnold et al on the torus dynamo. In this paper the eigenvalue for a Anosov filament may be expressed in terms of curvature $\kappa$ by $\kappa[\frac{-1\pm\sqrt{5}}{2}]$ while the Anosov space eigenvalues are given by $\frac{3\pm\sqrt{5}}{2}$. Note that when the Frenet curvature of the one dimensional dynamo plasma is $\kappa=-1$, yields $\frac{1\mp\sqrt{5}}{2}$ which is a simply slightly change on stretching and squeezing properties of dynamos with respect to the Arnold et al eigenvalue. Another type of one dimensional plasma by Fantoni and Tellez \cite{8} in the realm of electron plasmas in 2D. Here, as happens in general relativity the plasma undergoes a Coriolis force which is given by the presence of the curvilinear coordinates effects present in the Riemann-Christoffel symbol in the MHD dynamo equation. In their non-relativistic plasma limit, this geometry has been used by Fantoni and Tellez in the context of plasma physics. They have used a Flamm's paraboloid, which is a non-compact manifold which represents the spatial Schwarzschild black hole, to investigate one-component two-dimensional plasmas. Examples of dynamos in compact Anosov spaces, is the Moebius dynamo strip flow investigated by Shukurov et al \cite{9} in Perm compact dynamo torus experiment. Another example of dynamo plasma flows is given by examining the properties of Arnold Beltrami Childress [ABC] dynamo flows, whose stochastic properties have been investigated recently by Kleeorin et al \cite{10}. In this example, their geometrical properties are further investigated by considering ABC dynamo flows in curvilinear coordinates of the Riemannian space of twisted plasma tubes. Simulations with ABC dynamo flows have been obtained many years ago by Arnold and Korkina \cite{11}. Here one shall addopted a curvilinear coordinate system, to investigate geometrical and physical properties of ABC flows. This is given by the twisted system of coordinates used in general to investigate magnetic flux tubes \cite{12} in solar physics. In the second example one makes use of the stretched and squeezed metric given by a modification of Arnold, Zeldovich, Ruzmakin and Sokoloff [JETP 1981) in untwisted coordinates. In the first case introduction of complex coordinates makes computations easier. Stagnation points [Y Lau and J M Finn, Phys Fluids B 1993) in the dynamo resistive flow can be obtained for ABC flows in twisted tubes when constants $B=C$ vanish. This so called, strong stagnation point cannot support dynamo action, and only marginal dynamos are possible. In the presence of diffusion, slow dynamos are obtained. \newline
 Paper is organised as follows: Section II presents the mathematical formalism necessary to grasp the rest of the paper. In the next section the Frenet 2 plane magnetic fields across filamentary dynamo flows are presented. In section III the eigenvalues of the dynamo spectrum are obtained and the Anosov flow is obtained. ABC plasma dynamo example is addressed in section IV. Conclusions are presented in section V.
\newpage
\section{One dimensional plasmas and 2 planes magnetic fields}
In this paper a mathematical technique based on one dimensional plasma in the form of filaments in the Euclidean three dimensional space $\textbf{R}^{3}$. Here instead of complications of Riemannian spaces the one dimensional plasmas leave on a flat space where only non vanishing curvature is the scalar one given by Frenet frame of the filaments. This frame is given by the three basis vectors $[\textbf{t},\textbf{n},\textbf{b}]$,
This frame vectors obey the following evolution equations
\begin{equation}
\frac{d\textbf{t}}{ds}={\kappa}(s)\textbf{n}
\label{1}
\end{equation}
\begin{equation}
\frac{d\textbf{n}}{ds}=-{\kappa}(s)\textbf{t}+{\tau}\textbf{b}
\label{2}
\end{equation}
\begin{equation}
\frac{d\textbf{b}}{ds}=-{\tau}(s)\textbf{n}
\label{3}
\end{equation}
Here ${\kappa}$ and ${\tau}$ are Frenet curvature and torsion scalars. The magnetic self-induction equation is written as
\begin{equation}
\frac{{\partial}\textbf{B}}{{\partial}t}={\nabla}{\times}({\alpha}\textbf{B})+{\epsilon}{\Delta}\textbf{B}\label{4}
\end{equation}
where ${\alpha}=<\textbf{v}.{\nabla}{\times}\textbf{v}>$ represents the ${\alpha}$ helicity of the flow given by $\textbf{v}=v_{s}(s)\textbf{t}+v_{n}(s)\textbf{t}$. Note that though one considers here that the modulus of the flow the toroidal and normal components $v_{s}$ and $v_{n}$ depends on the toroidal coordinates. Even if one of these components is constant the flow is not necessarily laminar due to the dynamical unsteady nature of the frame vector $\textbf{t}$. This equation shall be expanded below, along the Frenet frame as
\begin{equation}
\textbf{B}(s,t)=B_{n}\textbf{n}+B_{b}\textbf{b}\label{5}
\end{equation}
Since in this section, one shall be considering the 2D case one shall assume from the beginning that the binormal component of the magnetic field $B_{b}$ shall vanish. Another simplification one shall addopt here is that the normal and binormal magnetic perturbations $B_{n}$ and $B_{b}$ shall be considered as constants. Some technical observations are in order now. The first is that this kind of Frenet frame used here, are called the isotropic Frenet frame, which considerers that even the frame is unsteady as here, in the sense that they depend upon time, the base vectors only depends upon the toroidal coordinate-s, and not other coordinates. Otherwise the Frenet frame is called anisotropic. The anisotropic may be more akin and suitable to turbulent phenomena but is much more involved, and shall be left to a next paper. Some simple use of the anisotropic Frenet frame can be found in simple Arnold like dynamos in reference \cite{10}. Let us start the MHD equations by the solenoidal divergence-free vector field by
\begin{equation}
{\nabla}.\textbf{B}=0 \label{6}
\end{equation}
as
\begin{equation}
{\partial}_{s}B_{b}-{\kappa}_{0}B_{n}=0 \label{7}
\end{equation}
Note that ${\kappa}_{0}={\tau}_{0}$ here represents the constant torsion ${\tau}_{0}$ equals the curvature ${\kappa}_{0}$ of the helical filamentary turbulence. By computing the relations
\begin{equation}
{\Delta}\textbf{t}=-{{\kappa}_{0}}^{2}\textbf{t}\label{8}
\end{equation}
\begin{equation}
{\Delta}\textbf{n}=-{{\kappa}_{0}}^{2}{\textbf{n}}\label{9}
\end{equation}
Since the 2D ${\alpha}$-dynamo is embedded in $\textbf{E}^{3}$, decomposition of the induction equation (\ref{4}), might be done along the three base vectors of the Frenet frame. Before starting the investigation of laminar dynamo flow with kinetic helicity, an important step is to derive the expression for ${\alpha}$. From the above expression for the kinetic helicity one obtains
\begin{equation}
{\alpha}=<\textbf{v}.{\nabla}{\times}\textbf{v}>=<\textbf{v}.{\omega}>\label{10}
\end{equation}
By replacing the definition of the flow above into this expression after some algebra yields
\begin{equation}
{\alpha}=-{\kappa}_{0} <{v_{n}}^{2}>\label{11}
\end{equation}
 where the symbol $<>$ indicates mean field objects, as given in mean field dynamos. In Anosov spaces, where the curvature is negative, the ${\alpha}$ effect is positive. By considering the Lyapunov chaotic behaviour of the magnetic field $|\textbf{B}|=B_{0}e^{{\lambda}t}$, the self induction equation is
\begin{equation}
{{\kappa}_{0}}B_{n}v_{s}=-{\kappa}_{0}B_{b}\label{12}
\end{equation}
\begin{equation}
{\gamma}B_{n}-{{\kappa}_{0}}B_{b}={\alpha}{\lambda}B_{n}-2{\beta}{{\kappa}_{0}}^{2}B_{n}\label{13}
\end{equation}
\begin{equation}
{\gamma}B_{b}-{{\kappa}_{0}}B_{n}v_{s}={\lambda}{\alpha}B_{n}-{\beta}{\kappa}_{0}B_{b}\label{14}
\end{equation}
Note that choice of 2-plane for the magnetic field polarised plane is transvected by the velocity of the Frenet curve.  This keeps some resemblance of the electric currents and the magnetic field which are orthogonal in general. This commonly happens in solid dynamos. The first dynamo expression, yields
\begin{equation}
B_{n}v_{s}=-B_{b}\label{15}
\end{equation}
which yields an expression for the rate at which normal component is transformed into the binormal one by the  velocity of one dimensional plasma flow. The remaining expressions can be collected in the form of a matrix to investigate the eigenvalue spectra
\begin{equation}
det[{\lambda}\textbf{I}-{D}_{\beta}]=0\label{16}
\end{equation}
Here $\textbf{I}$ represents the unit matrix
\vspace{1mm}
\begin{equation}$$\displaylines{\pmatrix{1&{0}\cr{0}&
1\cr}\cr}$$\label{17}
\end{equation}
where two-dimensional turbulent dynamo operator matrix $D_{\beta}$ can be written as
\begin{equation}
\vspace{1mm} $$\displaylines{\pmatrix{{\gamma}+2{\beta}{{\kappa}_{0}}^{2}-{\alpha}{\lambda}&{-{\kappa}_{0}}\cr
{-[{\alpha}{\lambda}+{{\kappa}_{0}}{v}_{s}]}&
{-[{\gamma}}+{\beta}{{\kappa}_{0}}^{2}]\cr}\cr}$$
\label{18}
\end{equation}
To simplify matters one shall choose the common practice in plasma physics of equipartition of magnetic field components in the form, $B_{n}=B_{b}$ which yields the following constraint on the flow, $v_{s}=-1$. Now within equipartition hypothesis, one shall address two cases of importance in dynamo physics: The first is the non turbulent ${\beta}=0$, laminar dynamo plasma case, which was also considered by Wang et al \cite{8} in the the cylindrical case without curvature or torsion. The second one when kinetic helicity vanishes while diffusive turbulence survives. In the first case the operator matrix ${D}$ reduces to
\begin{equation}
\vspace{1mm} $$\displaylines{\pmatrix{{\gamma}-{\alpha}{\lambda}&{-{\kappa}_{0}}\cr
{-[{\alpha}{\lambda}+{{\kappa}_{0}}{v}_{s}]}&
{-{\gamma}}\cr}\cr}$$\label{19}
\end{equation}
By taking the determinant of the matrix ${D}_{\alpha}$ one obtains the following algebraic second-order
\begin{equation}
{\gamma}^{2}+{\alpha}{\lambda}{{\gamma}}-[{\alpha}{\lambda}+{\kappa}_{0}]{\kappa}_{0}={0}\label{20}
\end{equation}
where one has considered that $v_{s}=-1$ and that the curvature is also very weak. Throghout the computations, the equipartition between curvature and torsion ${\kappa}_{0}={\tau}_{0}$ is assumed. By normalizing the curvature by ${\kappa}_{0}=1$ one obtains
\begin{equation}
{\gamma}_{\pm}= \frac{{\alpha}{\lambda}[-1\pm\sqrt{5}]}{2}\label{21}
\end{equation}
and since ${\alpha}{\lambda}=-1$ this expression finally reduces to
\begin{equation}
{\gamma}_{\mp}= \frac{[1\mp\sqrt{5}]}{2}\label{22}
\end{equation}
This eigenvalue characterizes Anosov spaces around torus manifold. Note that the first eigenvalue ${\gamma}_{-}\le{0}$, which indicates a non dynamo mode, while in the second case, ${\gamma}_{+}\ge{0}$, due to the expression
\begin{equation}
lim_{{\beta}\rightarrow{0}}\textbf{Re}{\gamma}(\beta)=0
\label{23}
\end{equation}
one obtains a fast dynamo mode. Here $\textbf{Re}$ represents the real part of the growth rate scalar ${\gamma}$. Since these eigenvalues represent the growth rate of the magnetic field, the plus sign represents the stretching of the filament while the minus sign represents the squeezing. Since stretching is fundamental for dynamo action , it is natural to find fast dynamos in the stretching mode. Note that if besides the non turbulent or laminar flow one also has a vanishing kinetic helicity only the marginal dynamo mode exists. In the second case, the turbulent dynamo operator ${D}_{\beta}$ becomes
\begin{equation}
\vspace{1mm}$$\displaylines{\pmatrix{{\gamma}+2{\beta}{{\kappa}_{0}}^{2}&{-{\kappa}_{0}}\cr
{-{{\kappa}_{0}}{v}_{s}}&{-[{\gamma}}+{\beta}{{\kappa}_{0}}^{4}]\cr}\cr}$$
\label{24}
\end{equation}
Before computing the eigenvalues for this matrix it is easy to show that the turbulent dynamo filaments reduces to a laminar oscillating dynamo mode, since in the limit of vanishing ${\beta}$ this matrix now reduces to
\begin{equation}
\vspace{1mm} $$\displaylines{\pmatrix{{\gamma}&{-{\kappa}_{0}}\cr
{-{\kappa}_{0}{v}_{s}}&
{-{\gamma}}\cr}\cr}$$
\label{25}
\end{equation}
which upon computation of its determinant yields
\begin{equation}
{\gamma}_{\pm}= \pm{i{\kappa}_{0}}\label{26}
\end{equation}
Since and ${\kappa}_{0}\in \textbf{R}$, this expression represents an oscillatory non Anosov dynamo. In general the eigenvalue problem yields the following spectra
\begin{equation}
{\gamma}_{\mp}= \frac{[1\mp\sqrt{5}]}{2}\label{27}
\end{equation}
which is still Anosov in general. Finally, let derive the expression for the discriminant
\begin{equation}
{\Delta}= b^{2}-4ac\label{28}
\end{equation}
corresponds to the algebraic second order equation,
\begin{equation}
a{\gamma}^{2}+b{\gamma}+c=0\label{29}
\end{equation}
Applying this case to the dynamo spectra above, one obtains in the laminar case, a relation for the kinematic and magnetic helicities, given by
\begin{equation}
{\Delta}= {\alpha}^{2}{\lambda}^{2}+4[{\alpha}{\lambda}+{\kappa}_{0}]{\kappa}_{0}=0\label{30}
\end{equation}
Solving this second order algebraic equation yields
\begin{equation}
{\alpha}{\lambda}= -2{\kappa}_{0}\label{31}
\end{equation}
which finally yields
\begin{equation}
{\gamma}_{\pm}= {\kappa}_{0}
\label{32}
\end{equation}
This shows that the fast dynamo condition ${\gamma}>0$, implies on this example that the filament possesses positive curvature.
\section{ABC flows in twisted Riemannian spaces}
Recently Kambe, Hattori and Zeitlin \cite{15} have investigated the rate of stretching of the flows by the Riemann curvature. They also showed that the ABC flow could be expressed in terms of the complex variables format as \begin{equation}
V_{ABC}= A(z)+B(x)+C(y)
\label{33}
\end{equation}
\begin{equation}
A= [(i,1,0)e^{iz}+(-i,1,0)e^{-iz}]
\label{34}
\end{equation}
\begin{equation}
B= [(0,i,1)e^{ix}+(0,-i,1)e^{-ix}]
\label{35}
\end{equation}
\begin{equation}
C=[(1,0,i)e^{iy}+(1,0,-i)e^{-iy}]
\label{36}
\end{equation}
Since the main task here is to transform the ABC flows given here in cartesian orthogonal $(x,y,z)$ into the coordinates of the twisted magnetic flux tubes given in the Riemann metric
\begin{equation}
{ds_{0}}^{2}= dr^{2}+ds^{2}+r^{2}d{\theta}^{2}
\label{37}
\end{equation}
where the coordinates of the tube are given by $[r,s,\theta]$ where $\theta[s]={\theta}_{0}-\int{{\tau}ds}$, is the twist coordinate, and since one already has the the velocity of ABC flows in cartesian coordinates, one just needs to write down the expression of coordinate transformation as
\begin{equation}
V_{x}={{\partial}_{s}x}{v}_{s}+{{\partial}_{r}}x{v}_{r}\label{38}
\end{equation}
\begin{equation}
V_{y}={\partial}_{s}y{v}_{s}+{\partial}_{r}y{v}_{r}\label{39}
\end{equation}
\begin{equation}
V_{z}={\partial}_{s}z{v}_{s}\label{40}
\end{equation}
along with the expression for change of coordinates
\begin{equation}
{x}=r cos{\theta}\label{41}
\end{equation}
\begin{equation}
{y}=r sin{\theta}\label{42}
\end{equation}
\begin{equation}
{z}\approx{s}\label{43}
\end{equation}
substituted into expressions above yields
\begin{equation}
V_{s}=B[e^{ircos{\theta}}+e^{-ircos{\theta}}]+C[e^{irsin{\theta}}+e^{-irsin{\theta}}]\label{44}
\end{equation}
\begin{equation}
V_{r}= \frac{csc{\theta}}{1+r}[iAm+Cin+rBi(e^{ircos{\theta}}-e^{-irsin{\theta}})]
\label{45}
\end{equation}
and taking the following form for the magnetic field
\begin{equation}
m=(e^{irs}-e^{-irs})\label{46}
\end{equation}
and
\begin{equation}
n=(e^{irsin{\theta}}-e^{-irsin{\theta}})\label{47}
\end{equation}
and
Close to the magnetic twisted flux tube where $r\approx{0}$, the radial flow becomes
\begin{equation}
V_{0}= [A{csc{\theta}}]s
\label{48}
\end{equation}
where one has made use of the approximation $sin s\approx{s}$ for the neighborhood of the origin of toroidal direction. Note then that last expression means that one obtained an oscillatory mode for ABC flow in twisted Riemannian space with growing amplitude. This is similar to what happens in La and Finn ABC  dynamo modes, nevertheless due to certain constraints here dynamo action is not possible. Substitution of approximation $sin[r{\theta}]\approx{r{\theta}}$ and similar approximations into the expression
\begin{equation}
{{\partial}_{t}}{\textbf{B}}=\textbf{B}.{\nabla}{\textbf{v}}\label{49}
\end{equation}
with the growth rate format of the magnetic field
\begin{equation}
\textbf{B}=\textbf{B}_{0}(\textbf{x})e^{{\gamma}t}\label{50}
\end{equation}
yields the following equations
\begin{equation}
\gamma{B}_{s}=[B_{s}-\frac{{{\tau}_{0}}^{-1}}{r^{2}}B_{\theta}]{\tau}_{0}V_{r}cos{\theta}
\label{51}
\end{equation}
\begin{equation}
\gamma\frac{B_{\theta}}{r^{2}}sin{\theta}=-[B_{s}-\frac{{{\tau}_{0}}^{-1}}{r^{2}}B_{\theta}]{\partial}_{s}V_{r}
cos{\theta}
\label{52}
\end{equation}
\begin{equation}
\gamma\frac{B_{\theta}}{r^{2}}cos{\theta}=[B_{s}-\frac{{{\tau}_{0}}^{-1}}{r^{2}}B_{\theta}]{\partial}_{s}V_{r}sin{\theta}
\label{53}
\end{equation}
multiplying the last two equations respectively by sin and cos one obtains after some trigometry,
\begin{equation}
\gamma\frac{B_{\theta}}{r^{2}}=0
\label{54}
\end{equation}
which shows that the local growth rate $\gamma[r,s]$ vanishes, unless the poloidal component of the magnetic field vanishes which would be a trivial case. Therefore the local growth rate vanishes which corresponds to a magnetic marginal dynamo. Substitution of this result into the equation for toroidal component $B_{s}$
reduces to
\begin{equation}
B_{s}=\frac{{{\tau}_{0}}^{-1}}{r^{2}}B_{\theta}\label{55}
\end{equation}
Just to complete the derivation, one may write the gradient of the radial component of the flow as eigenvalue spectra is determined. Dynamical effects of the stretching of magnetic fields by plasma flow \cite{12} may appear elsewhere. All the physical applications make the model presented here, useful in physical realistic situations and deserve further study.
\begin{equation}
{\partial}_{s}V_{r}=A[[1+tan^{2}{\theta}]{{\tau}_{0}}^{2}+cos s]
\label{56}
\end{equation}
where one has made use of the $B=C=0$ stagnation point condition. Actually in this case the toroidal flow vanishes. This certainly implies that no dynamo action is present. This gradient represents a growth rate of the radial velocity along the twisted magnetic axis. Let us now turn on diffusion, this situation shall turn on marginal dynamos into slow ones in the vicinity of stagnation points. This can be simply seen by recalling that the complete dynamo equation is given by
\begin{equation}
{{\partial}_{t}}{\textbf{B}}=\textbf{B}.{\nabla}{\textbf{v}}+\eta{\nabla}^{2}{\textbf{B}}\label{57}
\end{equation}
where $\eta$ is the constant diffusion constant. The simply presence of resistive term shows that the above expression where growth rate vanishes is not valid anymore and certainly slow dynamo modes might appear.

\section{Conclusions}
  It is shown that the presence of resistivity is  important to obtain at least slow ABC dynamos in Riemannian space, instead of marginal dynamos. Actually the presence of Riemann curvature, contributes to damp dynamo action instead of enhancing it. Lau and Finn \cite{14} have also discussed stagnation points in ABC flows. The action of Riemann curvature appears in the interaction of the Ricci tensor magnetic field and plasma resistivity. Stochastic problems in ABC dynamo flow were investigated by Kleeorin et al \cite{16}.  The simply presence of resistive plasmas are leads one from marginal modes to slow dynamo modes. A more detailed investigation of this point may appear elsewhere. Fast dynamos have been investigated in compact Riemannian manifolds, by Arnold et al \cite{1}. In this last case operator spectra in compact Riemannian spaces, have been determined. In this paper, fast kinematic are also investigated in the context of filaments in Anosov spaces \cite{10,11}. In this computation, Frenet holonomic frame is used, and the e
  \section{Acknowledgements}
  I am very much indebt to J-Luc Thiffeault, Dmitry Sokoloff, Yu Latushkin and Rafael Ruggiero for reading for helpful discussions on the subject of this work. I appreciate financial  supports from UERJ and CNPq.

\end{document}